\documentclass[journal]{IEEEtran}

\usepackage[T1]{fontenc}
\usepackage[utf8]{inputenc}

\usepackage{moresize}

\usepackage[
	backend=biber,
	style=numeric-comp,
	sorting=none,
	url=false
]{biblatex}
\AtBeginBibliography{\ssmall}

\AtEveryBibitem{
	\clearfield{urlyear}
	\clearfield{urlmonth}
}

\usepackage{csquotes}
\usepackage[english]{babel}

\usepackage[caption=false,font=normalsize,labelfont=sf,textfont=sf]{subfig}

\usepackage{graphicx}
\graphicspath{ {./} }

\usepackage[bookmarks,colorlinks]{hyperref}

\usepackage{amsmath}

\usepackage[edges]{forest}
\tikzset{
	section/.style = {align=center, text width=3cm},
	subsection/.style = {align=center, text width=3.5cm}
}

\title{A survey of deep learning audio generation methods}
\author{
	\IEEEauthorblockN{Matej Božić\IEEEauthorrefmark{1}, Marko Horvat\IEEEauthorrefmark{2}}
	\\
	\IEEEauthorblockA{
		Department of Applied Computing
		\\
		University of Zagreb, Faculty of Electrical Engineering and Computing, Zagreb, Croatia
		\\
		\IEEEauthorrefmark{1}matej.bozic@fer.hr
		\IEEEauthorrefmark{2}marko.horvat3@fer.hr
	}
}

\begin{document}

\maketitle

\begin{abstract}

This article presents a review of typical techniques used in three distinct aspects of deep learning model development for audio generation.
In the first part of the article, we provide an explanation of audio representations, beginning with the fundamental audio waveform. We then progress to the frequency domain, with an emphasis on the attributes of human hearing, and finally introduce a relatively recent development. 
The main part of the article focuses on explaining basic and extended deep learning architecture variants, along with their practical applications in the field of audio generation.
The following architectures are addressed: 1) Autoencoders 2) Generative adversarial networks 3) Normalizing flows 4) Transformer networks 5) Diffusion models.
Lastly, we will examine four distinct evaluation metrics that are commonly employed in audio generation.
This article aims to offer novice readers and beginners in the field a comprehensive understanding of the current state of the art in audio generation methods as well as relevant studies that can be explored for future research.

\end{abstract}


\section{Introduction}

\begin{figure}[!t]
	\scriptsize
	\centering
	\begin{forest}
		for tree={
			forked edges,
			grow'=0,
			draw,
			rounded corners
		},
		[Survey, rotate=90
			[\ref{sec:audio_features}. Audio features, for tree={fill=red!45}, section
				[Raw waveform, fill=red!30, subsection]
				[Mel-spectrogram, fill=red!30, subsection]
				[Neural codecs, fill=red!30, subsection]
			]
			[\ref{sec:architectures}. Architectures, for tree={fill=brown!45}, section
				[Auto-encoder, fill=brown!30, subsection]
				[Generative adversarial networks, fill=brown!30, subsection]
				[Normalizing flows, fill=brown!30, subsection]
				[Transformer networks, fill=brown!30, subsection]
				[Diffusion models, fill=brown!30, subsection]
			]
			[\ref{sec:eval_metrics}. Evaluation metrics, for tree={fill=orange!45}, section
				[Human evaluation, fill=orange!30, subsection]
				[Inception score, fill=orange!30, subsection]
				[Fréchet distance, fill=orange!30, subsection]
				[Kullback-Leibler divergence, fill=orange!30, subsection]
			]
		]
	\end{forest}
	\caption{The main sections of the survey.}
	\label{fig:survey_graph}
\end{figure}
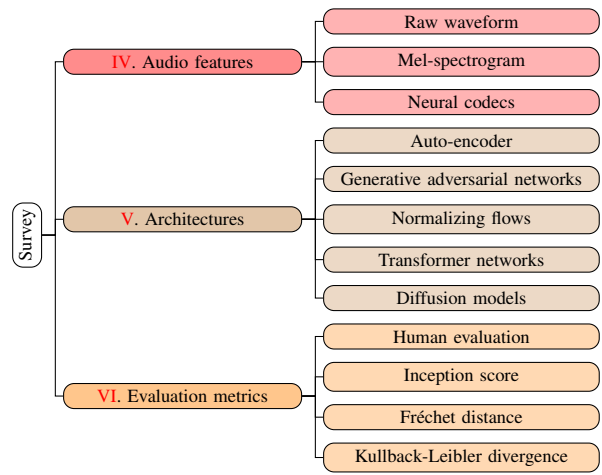

\IEEEPARstart{T}{he} trend towards deep learning in Computer Vision (CV) and Natural Language Processing (NLP) has also reached the field of audio generation \cite{peeters_DeepLearningAudio_2021}.
Deep learning has allowed us to move away from the complexity of hand-crafted features towards simple representations by letting the depth of the model create more complex mappings.
We define audio generation as any method whose outputs are audio and cannot be derived solely from the inputs.
Even though tasks such as text-to-speech involve translation from the text domain to the speech domain, there are many unknowns, such as the speaker's voice.
This means that the models have to invent or generate information for the translation to work.
There are many applications for audio generation.
We can create human-sounding voice assistants, generate ambient sounds for games or movies based on the current visual input, create various music samples to help music producers with ideas or composition, and much more.
The structure of the presented survey on deep learning audio generation methods is illustrated in Figure \ref{fig:survey_graph}.

This article will mainly focus on deep learning methods, as the field seems to be developing in this direction.
Nevertheless, section \ref{sec:background} will examine the development of audio generation methods over the years, starting around the 1970s.
We consider this section important because, just as deep learning methods have re-emerged, there may be a time when audio generation methods that are now obsolete become state-of-the-art again.
The goal is to take a broad but shallow look at the field of audio generation.
Some areas, such as text-to-speech, will be more heavily represented as they have received more attention, but an attempt has been made to include many different subfields.
This article does not attempt to present all possible methods but only introduces the reader to some of the popular methods in the field.
Each listing of works on a topic is sorted so that the most recent articles are at the end.

The article is structured as follows: section \ref{sec:related_work} presents previous work dealing with deep learning in audio, section \ref{sec:background} gives a brief overview of previous audio generation methods in text-to-speech and music generation, section \ref{sec:audio_features} deals with the two most prominent features and a recent advancement, section \ref{sec:architectures} discusses five deep learning architectures and some of their popular extensions, and finally, section \ref{sec:eval_metrics} looks at measuring the performance of generation models, some of which are specific to audio generation, while others are more generally applicable.

\section{Related work}
\label{sec:related_work}

In this section, we will mention some of the works that are good sources for further research in the field of audio generation.
Some of them investigate only a specific model architecture or sub-area, while others, like this work, show a broader view.

In \Textcite{zhao_ApplicationsDeepLearning_2019}, deep learning discriminative and generative architectures are discussed, along with their applications in speech and music synthesis.
The article covers discriminative neural networks such as Multi-Layer Perceptron (MLP), Convolutional Neural Networks (CNN), and Recurrent Neural Networks (RNN), as well as generative neural networks like Variational Autoencoders (VAE) and Deep Belief Networks (DBN).
They also describe generative adversarial networks (GAN), their flaws, and enhancement strategies (with Wasserstein GAN as a standout).
The study mainly focuses on speech generation and doesn't focus much on different hybrid models.

In contrast, \Textcite{purwins_DeepLearningAudio_2019} emphasizes other areas of modeling, including feature representations, loss functions, data, and evaluation methods.
It also investigates a variety of additional application fields, including enhancement as well as those outside of audio generation, such as source separation, audio classification, and tagging.
They describe various audio aspects that are not covered here, such as the mel frequency cepstral coefficients (MFCC) and the constant-Q spectrogram.
They do not cover as many architectures, but they do provide domain-specific datasets and evaluation methods.

Unlike previous works, \textcite{briot_DeepLearningTechniques_2019} attempts to comprehensively examine a specific field of audio generation.
This study considers five dimensions of music generation: objective, representation, architecture, challenge, and strategy.
It looks at a variety of representations, both domain-specific and more general.
Explains the fundamentals of music theory, including notes, rhythm, and chords. Introduces various previously established architectures such as MLP, VAE, RNN, CNN, and GAN, as well as some new ones like the Restricted Boltzmann Machine (RBM).
Finally, it discusses the many challenges of music generation and ways for overcoming them.
The work is quite extensive; however, some sections may benefit from a more detailed explanation.

\Textcite{huzaifah_DeepGenerativeModels_2020} is another work that explores the subject of music generation and includes music translation.
It discusses data representation, generative neural networks, and two popular DNN-based synthesizers.
It discusses the issue of long-term dependence and how conditioning might alleviate it.
Explains the autoregressive (AR) and normalized flow (NF) models, as well as VAE and GAN.

\Textcite{peeters_DeepLearningAudio_2021} provides an overview of deep learning techniques for audio.
It distinguishes architectures from meta-architectures.
The architectures include MLP, CNN, Temporal Convolutional Networks (TCN), and RNN, while the meta-architectures are Auto-Encoders (AE), VAE, GAN, Encoder/Decoder, Attention Mechanism, and Transformers.
Divides audio representations into three categories: time-frequency, waveform, and knowledge-driven.
Time-frequency representations include the Short-Time Fourier Transform (STFT), MFCC, Log-Mel-Spectrogram (LMS), and Constant-Q-Transform (CQT).
The article concludes with a list of applications for audio deep learning algorithms, including music content description, environmental sound description, and content processing. It also briefly discusses semi-supervised and self-supervised learning.

\Textcite{tan_SurveyNeuralSpeech_2021} provides a comprehensive overview of TTS methods, including history.
It explains the basic components of TTS systems, such as text analysis, acoustic models, and vocoders, and includes a list of models in each area. 
Finally, it discusses advanced methods for implementing TTS systems in certain use situations, such as Fast TTS, Low-Resource TTS, and Robust TTS.

\Textcite{shi_SurveyAudioSynthesis_2021} discusses TTS, music generation, audiovisual multi-modal processing, and datasets.
This effort differs from earlier ones in that it organizes relevant articles by category rather than explaining subjects in depth.

\Textcite{natsiou_AudioRepresentationsDeep_2021} is the closest work to this one.
It follows a similar structure, starting with input representations including raw waveforms, spectrograms, acoustic characteristics, embeddings, and symbolic representations, followed by conditioning representations used to guide audio synthesis.
Includes audio synthesis techniques such as AR, NF, GAN, and VAE.
The article concludes with the following evaluation methods: perceptual evaluation, number of statistically different bins, inception score, distance-based measurements, spectral convergence, and log likelihood.

\Textcite{latif_TransformersSpeechProcessing_2023} provides an overview of transformer architectures used in the field of speech processing.
The article provides a description of the transformer, a list of popular transformers for speech, and a literature review on its applications.

\Textcite{zhang_SurveyAudioDiffusion_2023} surveys TTS and speech enhancement, with a focus on diffusion models.
Although the emphasis is on diffusion models, they also discuss the stages of TTS, pioneering work, and specialized models for distinct speech enhancement tasks.

\Textcite{mehrish_ReviewDeepLearning_2023} conducted a comprehensive survey of deep learning techniques in speech processing.
It begins with speech features and traditional speech processing models.
It addresses the following deep learning architectures: RNN, CNN, Transformer, Conformer, Sequence-to-Sequence models (Seq2seq), Reinforcement learning, Graph neural networks (GNN), and diffusion probabilistic networks.
Explains supervised, unsupervised, semi-supervised, and self-directed speech representation learning.
Finally, it discusses a variety of speech processing tasks, including neural speech synthesis, speech-to-speech translation, speech enhancement, audio super resolution, as well transfer learning techniques.

\section{Background}
\label{sec:background}

The main purpose of this section is to show how audio generation has developed over the years up to this point. Since audio generation is a broad field that encompasses many different areas, such as text-to-speech synthesis, voice conversion, speech enhancement,... \cite{ling_DeepLearningAcoustic_2015}, we will only focus on two different areas of audio generation: text-to-speech synthesis and music generation. There is no particular reason for this choice, except that they are among the more popular ones. The trend we want to show is how domain-specific knowledge is shifting towards general-purpose methods and how feature engineering is turning into feature recognition.

\subsection{Text-to-Speech}

Text-to-speech (TTS) is a task with numerous applications, ranging from phone assistants to GPS navigators. The desire to construct a machine that can communicate with a human has historically fueled growth in this subject.
Conventional speech synthesis technologies include rule-based concatenative speech synthesis (CSS) and statistical parametric speech synthesis (SPSS) \cite{benesty_SpringerHandbookSpeech_2008}.
CSS and SPSS, which employ speech data, may be considered corpus-based speech synthesis approaches \cite{toda_SpeechParameterGeneration_2007}.

Until the late 1980s, the field was dominated by rule-based systems \cite{taylor_TexttoSpeechSynthesis_2009}.
They were heavily reliant on domain expertise such as phonological theory, necessitating the collaboration of many experts to develop a comprehensive rule set that would generate speech parameters.
There are numerous works like \textcite{liberman_MinimalRulesSynthesizing_1959,holmes_SpeechSynthesisRule_1964,coker_AutomaticSynthesisOrdinary_1973,klatt_StructurePhonologicalRule_1976}.

CSS methods try to achieve the naturalness and intelligibility of speech by combining chunks of recorded speech.
They can be divided into two categories: fixed inventory and unit-selection approaches \cite{benesty_SpringerHandbookSpeech_2008}.
Fixed inventory uses only one instance of each concatenative unit, which goes through signal processing before being combined into a spoken word.
An example of this might be \textcite{dixon_TerminalAnalogSynthesis_1968}, which uses the diphone method of segment assembly.
On the other hand, unit-selection employs a large number of concatenative units, which can result in a better match between adjacent units, potentially boosting speech quality.
There are two fundamental concepts: target cost and concatenation cost.
The target cost determines how well an element from a database of units fits the desired unit, whereas the concatenation cost indicates how well a pair of selected units combine.
The goal is to minimize both costs for the entire sequence of units; this is commonly done using a Viterbi search \cite{benesty_SpringerHandbookSpeech_2008}.
Although it is always possible to minimize costs, the resulting speech may still contain errors.
This can arise owing to a lack of units to choose from, an issue that can be mitigated by increasing the database size.
It sounds straightforward; however, doing so increases unit creation costs and search times due to the increased number of possible concatenations.
All of this requires CSS techniques to pick between speech quality and synthesis speed.
Works in this domain include \textcite{sagisaka_SpeechSynthesisRule_1988}, where they employ non-uniform synthesis units, and \textcite{hunt_UnitSelectionConcatenative_1996}, which treats units of a unit-selection database as states in a state transition network.

SPSS models speech parameters using statistical methods depending on the desired phoneme sequence.
This differs from CSS techniques in that we are not maintaining natural, unaltered speech but rather teaching the model how to recreate it.
In a typical SPSS system, this is done by first extracting parametric representations of speech and then modeling them using generative models, commonly by applying the maximum likelihood criterion \cite{zen_StatisticalParametricSpeech_2009}.
The primary advantage of SPSS over CSS is its ability to generalize to unknown data \cite{benesty_SpringerHandbookSpeech_2008}. This enables us to adjust the model to generate different voice characteristics \cite{taylor_TexttoSpeechSynthesis_2009}.
It also requires orders of magnitude less memory because we use model parameters instead of a speech database.
Although there are other SPSS techniques, the majority of research has centered on hidden Markov models (HMM) \cite{taylor_TexttoSpeechSynthesis_2009}.

Some HMM works include \textcite{toda_SpeechParameterGeneration_2007}, which considers not only the output probability of static and dynamic feature vectors but also the global variance (GV).
\Textcite{nakamura_IntegrationSpectralFeature_2014} directly models speech waveforms with a trajectory HMM.
\Textcite{yoshimura_SimultaneousModelingSpectrum_1999,tokuda_HMMbasedSpeechSynthesis_2002} use decision-tree-based context clustering to represent spectrum, pitch, and HMM state duration simultaneously.
Commonly used contexts include the current phoneme, preceding and succeeding phonemes, the position of the current syllable within the current word or phrase, etc. \cite{tokuda_SpeechSynthesisBased_2013}.

The notion that the human speech system has a layered structure in its transformation of the linguistic level to the waveform level has stimulated the adoption of deep neural network speech synthesis \cite{zen_StatisticalParametricSpeech_2013}.
\Textcite{burniston_HybridNeuralNetwork_1994} employs an artificial neural network alongside a rule-based method to model speech parameters.
\Textcite{ling_ModelingSpectralEnvelopes_2013} employs limited Boltzmann machines and deep belief networks to predict speech parameters for each HMM state.
Some other methods worth noting are multi-layer perceptron \cite{weijters_SpeechSynthesisArtificial_1993,lu_CombiningVectorSpace_2013,zen_StatisticalParametricSpeech_2013,fan_MultispeakerModelingSpeaker_2015,tokuday_DirectlyModelingSpeech_2015}, time-delay neural network \cite{karaali_SpeechSynthesisNeural_1998,karaali_TextToSpeechConversionNeural_1998}, long short-term memory \cite{fan_TTSSynthesisBidirectional_2014,zen_UnidirectionalLongShortterm_2015,li_MultiLanguageMultiSpeakerAcoustic_2016,tokuda_DirectlyModelingVoiced_2016}, gated recurrent unit \cite{wang_GatingRecurrentMixture_2016}, attention-based recurrent network \cite{wang_FirstStepEndtoEnd_2016}, and mixture density network \cite{zen_DeepMixtureDensity_2014,wang_GatingRecurrentMixture_2016}.

The TTS system consists of four major components: the first converts text to a linguistic representation, the second determines the duration of each speech segment, the third converts the linguistic and timing representations into speech parameters, and the fourth is the vocoder, which generates the speech waveform based on the speech parameters \cite{karaali_SpeechSynthesisNeural_1998}.
The majority of the works we presented focused on converting the linguistic representation into speech parameters, but there are also models focusing on, for example, grapheme-to-phoneme conversion \cite{rao_GraphemetophonemeConversionUsing_2015,yao_SequencetoSequenceNeuralNet_2015} to allow TTS without knowledge of linguistic features.
Examples of vocoders include MLSA \cite{imai_CepstralAnalysisSynthesis_1983}, STRAIGHT \cite{kawahara_RestructuringSpeechRepresentations_1999}, and
Vocaine \cite{agiomyrgiannakis_VocaineVocoderApplications_2015}.
Finally, there have also been attempts to construct a fully end-to-end system, which means integrating text analysis and acoustic modeling into a single model \cite{wang_FirstStepEndtoEnd_2016}.

\subsection{Music generation}

Music has been a part of human life long before the invention of the electronic computer, and people have developed many guidelines for how beautifully sounded music should be made.
For this reason alone, the discipline of music generation has placed a heavy emphasis on rule-based systems that use music theory to create logical rules.
Unlike text, musical vocabulary is rather tiny, consisting of at most several hundred discrete note symbols \cite{lavrenko_PolyphonicMusicModeling_2003}.
Music creation is classified into six categories: grammars, knowledge-based, markov chains, artificial neural networks, evolutionary methods, and self-similarity \cite{fernandez_AIMethodsAlgorithmic_2013}.
Specific methods include discrete nonlinear maps \cite{pressing_NonlinearMapsGenerators_1988, dodge_ComputerMusicSynthesis_1985}, rule-based \cite{giomi_ComputationalGenerationStudy_1991, ames_CyberneticComposerOverview_1992}, genetic algorithm \cite{biles_GenJamGeneticAlgorithm_1994, moroni_VoxPopuliInteractive_2000, chen_CreatingMelodiesEvolving_2001, delapuente_AutomaticCompositionMusic_2002}, recurrent neural network \cite{mozer_NeuralNetworkMusic_1994, chen_CreatingMelodiesEvolving_2001}, long short-term memory \cite{eck_FirstLookMusic_2002, chu_SongPIMusically_2016, huang_DeepLearningMusic_2016, mogren_CRNNGANContinuousRecurrent_2016}, markov chain \cite{dodge_ComputerMusicSynthesis_1985, jones_CompositionalApplicationsStochastic_1981}, context-free grammars \cite{jones_CompositionalApplicationsStochastic_1981, delapuente_AutomaticCompositionMusic_2002}, context-sensitive grammars \cite{kohonen_SelflearningMusicalGrammar_1989, kohonen_NonheuristicAutomaticComposing_1991}, cellular automaton \cite{miranda_GranularSynthesisSounds_1995}, random fields \cite{lavrenko_PolyphonicMusicModeling_2003}, L-systems \cite{worth_GrowingMusicMusical_2005}, knowledge base \cite{chan_ImprovingAlgorithmicMusic_2006}, and restricted Boltzmann machines \cite{boulanger-lewandowski_ModelingTemporalDependencies_2012}.
Unlike language, music employs a significantly smaller number of acoustic features.
These include MIDI representation \cite{moroni_VoxPopuliInteractive_2000,huang_DeepLearningMusic_2016}, encoded sheet music \cite{chen_CreatingMelodiesEvolving_2001}, binary vector of an octave \cite{lavrenko_PolyphonicMusicModeling_2003}, and piano roll \cite{boulanger-lewandowski_ModelingTemporalDependencies_2012,huang_DeepLearningMusic_2016}.

\section{Audio features}
\label{sec:audio_features}

Even though there have been numerous audio features used throughout the history of audio generation.
Here we will describe the two most popular features, the raw waveform and the log-mel spectrogram, but also mention features that have recently gained traction.
Keep in mind that there are too many features to describe them all, especially if we take into account the many hand-crafted features that were created before the rise of deep learning methods.

\subsection{Raw waveform}
\label{ssec:waveform}

The term "raw audio" typically refers to a waveform recorded using pulse code modulation (PCM) \cite{natsiou_AudioRepresentationsDeep_2021}. In PCM, a continuous waveform is sampled at uniform intervals, known as the sampling frequency.
According to the sampling principle, if a signal is sampled at regular intervals at a rate slightly higher than twice the highest signal frequency, then it will contain all of the original signal information \cite{black_PulseCodeModulation_1947}.
The average sample frequency for audio applications is 44.1 kHz \cite{natsiou_AudioRepresentationsDeep_2021}, hence we cannot hold frequencies equal to or greater than 22.05 kHz.
Computers cannot store real numbers with absolute precision; thus, each sample value is approximated by assigning it an element from a set of finite values, a technique known as quantization \cite{natsiou_AudioRepresentationsDeep_2021}.
The most common quantization levels are kept in 8 bits (256 levels), 16 bits (65536 levels), and 24 bits (16.8 million levels) \cite{natsiou_AudioRepresentationsDeep_2021}.

The advantage of using raw audio waveforms is that they can be easily transformed into actual sound. In certain tasks, the disadvantages appear to outweigh the benefits, as raw waveforms are still not universally used.
The issue is that raw audio synthesis at higher bit rates becomes problematic due to the sheer amount of states involved \cite{verma_GenerativeModelRaw_2021}. For example, 24-bit audio signals have more than 16 million states.
High sampling rates create exceptionally long sequences, making raw audio synthesis more challenging \cite{wang_NeuralCodecLanguage_2023}.
\(\mu\)-law is frequently employed in speech generative models like WaveNet \cite{oord_WaveNetGenerativeModel_2016} to compress integer values and sequence length. The method can quantize each timestep to 256 values and reconstruct high-quality audio \cite{wang_NeuralCodecLanguage_2023}.
According to \textcite{dieleman_ChallengeRealisticMusic_2018}, increased bit depth representation can lead to models learning undesirable aspects, such as the calm background of the surroundings.
It should be emphasized that this issue was only observed in older publications and is not discussed in current ones.

The most common models that use raw waveforms as their representation of choice are text-to-speech models called vocoders. In section \ref{sec:background}, we mentioned vocoders, which are used to translate mid-term representations, such as mel-spectrograms, to raw audio waveforms. Examples include WaveNet \cite{oord_WaveNetGenerativeModel_2016}, SampleRNN \cite{mehri_SampleRNNUnconditionalEndtoEnd_2017}, and Deep Voice 3 \cite{ping_DeepVoiceScaling_2018}.

\subsection{Mel-spectrogram}
\label{ssec:mel}

Before we can talk about mel-spectrograms, we must first understand the Short-Time Fourier Transform (STFT).
To represent audio frequencies, we use a Discrete-Fourier Transform (DFT), which transforms the original waveform into a sum of weighted complex exponentials \cite{prandoni_SignalProcessingCommunications_2008}.
The problem emerges when we attempt to analyze complex audio signals; because the content of most audio signals changes over time, we can't use DFT to figure out how frequencies change.
Instead, we use STFT to apply DFT to overlapping sections of the audio waveform \cite{natsiou_AudioRepresentationsDeep_2021}.
Most techniques that use the STFT to represent audio consider just its amplitude \cite{peeters_DeepLearningAudio_2021}, which results in a lossy representation.
By removing the phase of the STFT, we can arrange it in a time/frequency visual, creating a spectrogram.

A mel-spectrogram compresses the STFT in the frequency axis by projecting it onto a scale known as the mel-scale \cite{nistal_ComparingRepresentationsAudio_2021}.
The mel-scale divides the frequency range into a set of mel-frequency bands, with higher frequencies having lower resolution and lower frequencies having higher resolution \cite{mehrish_ReviewDeepLearning_2023}.
The scale was inspired by the non-linear frequency perception of human hearing \cite{mehrish_ReviewDeepLearning_2023}.
Applying the logarithm to the amplitude results in the log-mel-spectrogram \cite{peeters_DeepLearningAudio_2021}.
Finally, using the discrete cosine transform yields the mel frequency cepstral coefficients (MFCC) \cite{purwins_DeepLearningAudio_2019}.
MFCC is a popular representation in speech applications \cite{benesty_SpringerHandbookSpeech_2008}, but it was shown to be unnecessary with deep learning models \cite{purwins_DeepLearningAudio_2019, peeters_DeepLearningAudio_2021}.

While representations such as the STFT and raw waveform are invertible, the spectrogram is not, so we must use some approach to approximate the missing values.
The algorithms used for these were already mentioned in the previous section. 
In addition to neural-based vocoders, other algorithms include Griffin-Lim \cite{griffin_SignalEstimationModified_1984}, gradient-based inversion \cite{decorsiere_InversionAuditorySpectrograms_2015}, single-pass spectrogram inversion (SPSI) \cite{beauregard_SinglePassSpectrogram_2015}, and phase gradient heap integration (PGHI) \cite{prusa_NoniterativeMethodReconstruction_2017}.
Mel-spectrograms have been frequently utilized as intermediate features in text-to-speech pipelines \cite{shi_SurveyAudioSynthesis_2021}.
Tacotron 1/2 \cite{wang_TacotronEndtoEndSpeech_2017,shen_NaturalTTSSynthesis_2018} and FastSpeech 1/2 \cite{ren_FastSpeechFastRobust_2019,ren_FastSpeechFastHighQuality_2022} are examples of such models.
To better illustrate the compression of the mel-spectrogram, take, for example, a 5-minute video sampled at 44.1 kHz.
With 16-bit depth, our raw waveform will take up \({\approx} 25 \text{MB}\), while the mel-spectrogram with a common configuration\footnote{Based on a small sample of articles observed using mel-spectrograms} of 80 bins, 256 hop size, 1024 window size, and 1024 points of Fourier transform takes up \({\approx} 8 \text{MB}\) at the same bit depth.
In a \citeyear{choi_ComparisonAudioSignal_2018} article, it was proven that mel-spectrogram is preferable over STFT because it achieves the same performance while having a more compact representation \cite{choi_ComparisonAudioSignal_2018}.
Given the field's progress, it should be emphasized that only recurrent and convolutional models were examined in the article.
Another advantage of the mel-spectrogram, and spectrograms in general, is that they can be displayed as images.
Because it ignores phase information, it can be shown with one dimension being frequency and the other being time.
This is useful since images have been widely employed in computer vision tasks, allowing us to borrow models for usage in the audio domain.
There is a concern as spectrograms aren't the same as images due to the different meaning of the axis.
This does not appear to have a substantial effect since many works implement mel-spectrograms in their convolutional models \cite{peeters_DeepLearningAudio_2021}.

\subsection{Neural codecs}
\label{ssec:neural_codec}

An audio codec is a signal processing technique that compresses an audio signal into discrete codes before using those codes to reconstruct the audio signal, which is not always possible with complete accuracy.
A typical audio codec system consists of three components: an encoder, a quantizer, and a decoder. The function of each component is explained in section \ref{ssec:auto_encoder}.
The goal of an audio codec is to use as little information as possible to store or transmit an audio signal while ensuring that the decoded audio quality is not significantly reduced by eliminating redundant or irrelevant information from the audio signal.
Traditionally, this is accomplished by changing the signal and trading off the quality of specific signal components that are less likely to influence the quality \cite{defossez_HighFidelityNeural_2022}.
Audio codecs have been utilized for a wide range of applications, including mobile and internet communication.
There are numerous types of audio codecs; some are utilized in real-time applications like streaming, while others may be used for audio production.
Whereas in streaming, latency is a larger concern, which means sacrificing quality for speed, in production, we want to retain as much detail as possible while maintaining a compact representation.

Audio codecs can be separated into two categories: waveform codecs and parametric codecs.
Waveform codecs make little to no assumptions about the nature of audio, allowing them to work with any audio signal.
This universality makes them well-suited for creating high-quality audio at low compression, but they tend to produce artifacts when operating at high compression \cite{zeghidour_SoundStreamEndtoEndNeural_2022}.
Furthermore, because they do not operate well in high compression, they tend to increase storage and transmission costs.
In contrast to waveform codecs, parametric codecs make assumptions about the source audio being encoded and introduce strong priors in the form of a parametric model that characterizes the audio synthesis process.
The goal is not to achieve a faithful reconstruction on a sample-by-sample basis but rather to generate audio that is perceptually comparable to the original \cite{zeghidour_SoundStreamEndtoEndNeural_2022}.
Parametric codecs offer great compression but suffer from low decoded audio quality and noise susceptibility \cite{ai_APCodecNeuralAudio_2024}.

On the way to the neural codec, we first encountered hybrid codecs, which substituted some parametric codec modules with neural networks.
This type of codec improves performance by leveraging neural networks' adaptability.
Following that came vocoder-based approaches, which could leverage previously introduced neural vocoders to reconstruct audio signals by conditioning them on parametric coder codes or quantized acoustic features.
However, their performance and compression were still dependent on the handcrafted features received at the input \cite{wu_AudioDecOpensourceStreaming_2023}.
The observation that separating models into modules prevents them from functioning effectively has inspired end-to-end auto-encoders (E2E AE) that accept raw waveforms as input and output.
A standard E2E AE is made up of four basic components: an encoder, a projector, a quantizer, and a decoder \cite{wu_AudioDecOpensourceStreaming_2023}.
The basic use case is to take the raw waveform and use the encoder to construct a representation with reduced temporal resolution, which is then projected into a multidimensional space by the projector component.
To make the representations suitable for transmission and storage, we further quantize the projections into codes.
These codes make up a lookup table, which is used at the other end by the decoder to transform the quantized representations back to a raw waveform.
\Textcite{shen_NaturalSpeechLatentDiffusion_2023} defines the neural codec as a kind of neural network model that converts audio waveform into compact representations with a codec encoder and reconstructs audio waveform from these representations with a codec decoder.
The core idea is to use the audio codec to compress the speech or sound into a set of discrete tokens, and then the generation model is used to generate these tokens \cite{yang_HiFiCodecGroupresidualVector_2023}.
They have been shown to allow for cross-modal tasks \cite{wang_VioLAUnifiedCodec_2023}.

Neural codec methods include SoundStream \cite{zeghidour_SoundStreamEndtoEndNeural_2022}, EnCodec \cite{defossez_HighFidelityNeural_2022}, HiFi-Codec \cite{yang_HiFiCodecGroupresidualVector_2023}, AudioDec \cite{wu_AudioDecOpensourceStreaming_2023}, and APCodec \cite{ai_APCodecNeuralAudio_2024}.
All the said methods use residual vector quantization (RVQ), while HiFi-Codec also introduced an extension called group-RVQ. VQ methods will be talked about in section \ref{ssec:auto_encoder}.
SoundStream \cite{zeghidour_SoundStreamEndtoEndNeural_2022} is used by AudioLM, \cite{borsos_AudioLMLanguageModeling_2023}, MusicLM \cite{agostinelli_MusicLMGeneratingMusic_2023}, SingSong \cite{donahue_SingSongGeneratingMusical_2023} and SoundStorm \cite{borsos_SoundStormEfficientParallel_2023}, while
EnCodec \cite{defossez_HighFidelityNeural_2022} is used by VALL-E \cite{wang_NeuralCodecLanguage_2023}, VALL-E X \cite{zhang_SpeakForeignLanguages_2023}, Speech-X \cite{wang_SpeechXNeuralCodec_2023a}, and VioLA \cite{wang_VioLAUnifiedCodec_2023}.

Finally, despite the fact that neural codec approaches are relatively new, they have not been without criticism.
\Textcite{shen_NaturalSpeechLatentDiffusion_2023} noted that although RVQ can achieve acceptable reconstruction quality and low bitrate, they are meant for compression and transmission; therefore, they may not be suited as intermediate representations for audio production jobs.
This is because the sequence of discrete tokens created by RVQ can be very long, approximately \(N\) times longer when \(N\) residual quantifiers are utilized.
Because language models cannot handle extremely long sequences, we will encounter inaccurate predictions of discrete tokens, resulting in word skipping, word repetition, or speech collapse issues while attempting to reconstruct the speech waveform from these tokens.

\section{Architectures}
\label{sec:architectures}

As the models become more advanced, they start utilizing many different architectures in unison, making it impossible to categorize them efficiently. Therefore, each subsection will contain models that fit into many subsections but have been divided up in the way the author thought made the most sense. Unlike the audio features, there are many different architectures. Here we will mention the architectures that have been most commonly used in the field of audio generation.

\subsection{Auto-encoders}
\label{ssec:auto_encoder}

The majority of this section was taken from works by \textcite{goodfellow_DeepLearning_2016, kingma_IntroductionVariationalAutoencoders_2019}.

The auto-encoder's objective is to duplicate the input into the output.
It consists of two parts: an encoder and a decoder.
The intersection of the two components depicts a code that attempts to represent both the input and, by extension, the output.
The encoder receives input data and changes it into a code that the decoder then uses to approximate the original input.
If we allowed arbitrary values in the encoder and decoder, we would obtain no meaningful code because it would simply simulate an identity function.
To obtain meaningful code, we constrain both the encoder and decoder, preventing them from just passing data through.
We can accomplish this by, for example, restricting the dimensionality of the values in the model.
The auto-encoder has the advantage of not requiring labeled data because it merely seeks to reconstruct the input, allowing for unsupervised learning.
\Textcite{lee_AudioFeatureGeneration_2019} demonstrates a basic use case for a simple auto-encoder, feature extraction.
While VITS \cite{kim_ConditionalVariationalAutoencoder_2021} connects two text-to-speech modules using VAE, enabling end-to-end learning in an adversarial setting.
Figure \ref{fig:architectures:autoencoder} depicts a simple auto-encoder setup using generic encoder and decoder components.

As this article focuses on generation, we will now introduce one of the most popular forms of the auto-encoder, the Variational Auto-Encoder (VAE) \cite{peeters_DeepLearningAudio_2021}.
VAE has been proposed to enable us to employ auto-encoders as generative models \cite{natsiou_AudioRepresentationsDeep_2021}.
The VAE components can be considered as a combination of two separately parameterized models, the recognition model and the generative model.
The VAE's success was mostly due to the choice of the Kullback-Leibler (KL) divergence as the loss function \cite{natsiou_AudioRepresentationsDeep_2021}. KL will also be described in section \ref{sec:eval_metrics}.
Unlike the auto-encoder, the VAE learns the parameters of a probability distribution rather than a compressed representation of the data \cite{huzaifah_DeepGenerativeModels_2020}.
Modeling the probability distribution allows us to sample from the learned data distribution.
The Gaussian distribution is typically used for its generality \cite{briot_DeepLearningTechniques_2019}.
MusicVAE \cite{roberts_HierarchicalLatentVector_2018} uses a hierarchical decoder in a recurrent VAE to avoid posterior collapse.
BUTTER \cite{zhang_BUTTERRepresentationLearning_2020} creates a unified multi-model representation learning model using VAEs.
Music FaderNets \cite{tan_MusicFaderNetsControllable_2020} introduces a Guassian Mixture VAE.
RAVE \cite{caillon_RAVEVariationalAutoencoder_2021} employs multi-stage training, initially with representation learning and then with adversarial fine-tuning.
Tango \cite{ghosal_TexttoAudioGenerationUsing_2023} and Make-An-Audio 2 \cite{huang_MakeAnAudioTemporalEnhancedTexttoAudio_2023} generate mel-spectrograms by using VAE in a diffusion model.

Vector-Quantized VAE (VQ-VAE) is an extension of VAE that places the latent representation in a discrete latent space.
VQ-VAE changes the auto-encoder structure by introducing a new component called the codebook.
The most significant change happens between the encoder and decoder, where the encoder's output is used in a nearest neighbor lookup utilizing the codebook.
In other words, the continuous value received from the encoder is quantized and mapped onto a discrete latent space that will be received by the decoder.
VQ-VAE replaces the KL divergence loss with negative log likelihood, codebook, and commitment losses.
One possible issue with the VQ-VAE is codebook collapse.
This occurs when the model stops using a piece of the codebook, indicating that it is no longer at full capacity.
It can result in decreased likelihoods and inadequate reconstruction \cite{dieleman_ChallengeRealisticMusic_2018}.
\Textcite{dieleman_ChallengeRealisticMusic_2018} proposes the argmax auto-encoder as an alternative to VQ-VAE for music generation.
MelGAN \cite{kumar_MelGANGenerativeAdversarial_2019}, VQVAE \cite{tjandra_VQVAEUnsupervisedUnit_2019}, Jukebox \cite{dhariwal_JukeboxGenerativeModel_2020} with Hierarchical VQ-VAE, DiscreTalk \cite{hayashi_DiscreTalkTexttoSpeechMachine_2020}, FIGARO \cite{rutte_FIGAROControllableMusic_2022}, Diffsound \cite{yang_DiffsoundDiscreteDiffusion_2023}, and Im2Wav \cite{sheffer_HearYourTrue_2023} use VQ-VAE to compress the input to a lower-dimensional space.
While Dance2Music-GAN \cite{zhu_QuantizedGANComplex_2022}, SpeechT5 \cite{ao_SpeechT5UnifiedModalEncoderDecoder_2022}, VQTTS \cite{du_VQTTSHighFidelityTexttoSpeech_2022}, and DelightfulTTS 2 \cite{liu_DelightfulTTSEndtoEndSpeech_2022} only use the vector quantization.

Residual Vector Quantization (RVQ) improves VAE by computing the residual after quantization and further quantizing it using a second codebook, a third, and so on \cite{defossez_HighFidelityNeural_2022}.
In other words, RVQ cascades \(N\) layers of VQ where unquantized input vector is passed through a first VQ and quantization residuals are computed, then those residuals are iteratively quantized by an additional sequence of \(N - 1\) vector quantizers \cite{zeghidour_SoundStreamEndtoEndNeural_2022}.
Section \ref{ssec:neural_codec} provides a list of models that employ RVQ.

\subsection{Generative adversarial networks}

\begin{figure*}[!t]
	\centering
	\subfloat[][Auto-encoder]{
		\includegraphics[width=0.7\columnwidth]{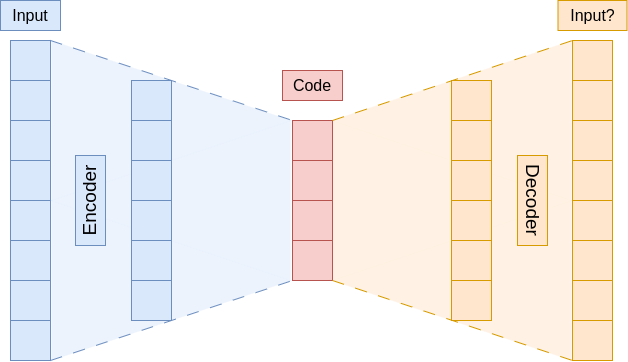}
		\label{fig:architectures:autoencoder}
	}
	\hfil
	\subfloat[][Generative adversarial network]{
		\includegraphics[width=0.9\columnwidth]{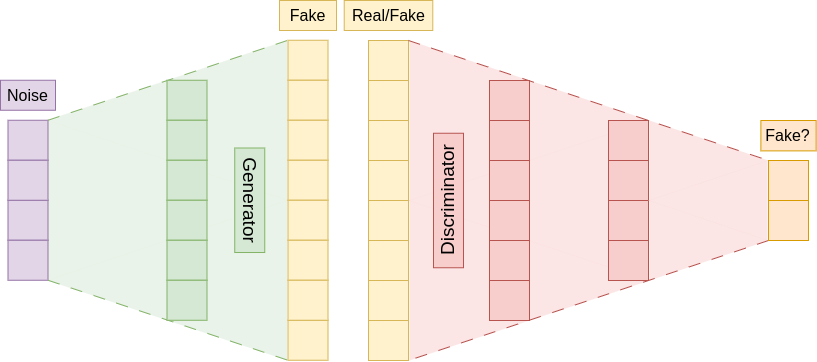}
		\label{fig:architectures:gan}
	}
	\caption{Two deep learning architectures that appear to have little in common until we look closer. The generator mimics the auto-encoder's decoder, whereas the discriminator resembles the encoder.}
	\label{fig:architectures}
\end{figure*}

Another prominent generating architecture is generative adversarial networks (GAN).
It is made up of two models that serve different purposes: the generator and the discriminator.
The generator's function is to convert a random input vector to a data sample.
The random vector is usually smaller because the generator mimics the decoder part of the auto-encoder \cite{peeters_DeepLearningAudio_2021}.
Unlike VAE, which imposes a distribution to generate realistic data, GAN utilizes a second network called the discriminator \cite{peeters_DeepLearningAudio_2021}.
It takes the generator's output or a sample from the dataset and attempts to classify it as either real or fake.
The generator is penalized based on the discriminator's ability to tell the difference between real and fake.
The opposite is also true: if the discriminator is unable to distinguish between the generator and the actual data points, it is penalized as well.
In other words, the two neural networks face off in a two-player minimax game.
According to \cite{torres-reyes_AudioEnhancementSynthesis_2019}, the ideal outcome for network training is for the discriminator to be 50\% certain whether the input is real or bogus.
In practice, we train the generator through the discriminator by reducing the probability that the sample is fake, while the discriminator does the opposite for fake data and the same for real data.
Figure \ref{fig:architectures:gan} illustrates the generator taking a random vector input and the discriminator attempting to distinguish between real and fake samples.

This basic setup allows us to generate samples resembling those in the dataset, but it doesn't let us condition the generation.
In other words, the random vector used in the generator does not match the semantic features of the data \cite{huzaifah_DeepGenerativeModels_2020}.
Many datasets contain additional information about each sample, such as the type of object in an image.
It would be beneficial if we could use the additional information to condition the generator and generate from a subset of the learned outputs.
Conditional GAN (cGAN) induces additional structure by including additional information into the generator and discriminator inputs.
The generator adds the additional information to the random vector, whereas the discriminator adds it to the data to discriminate.
Some of the works that utilize cGAN are MidiNet \cite{yang_MidiNetConvolutionalGenerative_2017}, \textcite{michelsanti_ConditionalGenerativeAdversarial_2017}, \textcite{chen_DeepCrossModalAudioVisual_2017}, \textcite{neekhara_ExpeditingTTSSynthesis_2019}, and V2RA-GAN \cite{liu_EndtoEndVisualtoRawAudioGeneration_2022}.

Common issues with GAN include mode collapse, unstable training, and a lack of an evaluation metric \cite{zhao_ApplicationsDeepLearning_2019}.
Mode collapse occurs when the generator focuses exclusively on a few outputs that can trick the discriminator into thinking they are real.
Even if the generator meets the discriminator requirements, we cannot use it to produce more than a few examples.
This might happen because the discriminator is unable to force the generator to be diverse \cite{torres-reyes_AudioEnhancementSynthesis_2019}.
The Wasserstein GAN (WGAN) is a well-known variant for addressing this problem. WGAN shifts the discriminator's job from distinguishing between real and forged data to computing the Wasserstein distance, commonly known as the Earth Mover's distance.
In addition, a modification to aid WGAN convergence has been proposed; it uses a gradient penalty rather than weight clipping and is known as WGAN-GP.
WGAN was used by MuseGAN \cite{dong_MuseGANMultitrackSequential_2018}, WaveGAN \cite{donahue_AdversarialAudioSynthesis_2019}, TiFGAN \cite{marafioti_AdversarialGenerationTimeFrequency_2019}, and Catch-A-Waveform \cite{greshler_CatchAWaveformLearningGenerate_2021}.

Another popular modification to the GAN architecture is the use of deep convolutional networks known as deep convolutional GANs (DCGAN).
Unlike WGAN, DCGAN only requires a change to the model architecture, rather than the entire training procedure, for both the generator and discriminator.
It aims to provide a stable learning environment in an unsupervised setting by applying a set of architectural constraints \cite{torres-reyes_AudioEnhancementSynthesis_2019}.
DGAN is used in works such as MidiNet \cite{yang_MidiNetConvolutionalGenerative_2017}, WaveGAN \cite{donahue_AdversarialAudioSynthesis_2019}, and TiFGAN \cite{marafioti_AdversarialGenerationTimeFrequency_2019}.

Furthermore, it is worth noting a simple GAN extension designed to address the issue of vanishing gradients while simultaneously improving training stability.
Least squares GAN (LSGAN) improves the quality of generated samples by altering the discriminator's loss function.
Unlike the regular GAN, LSGAN penalizes correctly classified samples much more, pulling them toward the decision boundary, which allows LSGAN to generate samples that are closer to the real data \cite{mao_LeastSquaresGenerative_2017}.
Papers using LSGAN include SEGAN \cite{pascual_SEGANSpeechEnhancement_2017}, \textcite{yamamoto_ProbabilityDensityDistillation_2019}, HiFi-GAN \cite{kong_HiFiGANGenerativeAdversarial_2020}, Parallel WaveGAN \cite{yamamoto_ParallelWaveganFast_2020}, Fre-GAN \cite{kim_FreGANAdversarialFrequencyconsistent_2021}, VITS \cite{kim_ConditionalVariationalAutoencoder_2021} and V2RA-GAN \cite{liu_EndtoEndVisualtoRawAudioGeneration_2022}.

There are many more modifications to GAN we haven't mentioned, like the Cycle GAN \cite{hao_CMCGANUniformFramework_2018} or the Boundary-Equilibrium GAN \cite{liu_UnconditionalAudioGeneration_2020}, as we tried to showcase the most prevalent modifications in the field of audio generation.
Works like MelGAN \cite{kumar_MelGANGenerativeAdversarial_2019}, GAAE \cite{haque_HighFidelityAudioGeneration_2020}, GGAN \cite{haque_GuidedGenerativeAdversarial_2021}, SEANet \cite{tagliasacchi_SEANetMultimodalSpeech_2020}, EATS \cite{donahue_EndtoEndAdversarialTexttoSpeech_2021}, Dance2Music-GAN \cite{zhu_QuantizedGANComplex_2022} and Musika \cite{pasini_MusikaFastInfinite_2022}
use yet another type of loss called the hinge loss.

Finally, we'd like to mention works that were challenging to categorize. They are GANSynth \cite{engel_GANSynthAdversarialNeural_2019}, GAN-TTS \cite{binkowski_HighFidelitySpeech_2019}, RegNet \cite{chen_GeneratingVisuallyAligned_2020}, Audeo \cite{su_AudeoAudioGeneration_2020}, Multi-Band MelGAN \cite{yang_MultiBandMelganFaster_2021}, Multi-Singer \cite{huang_MultiSingerFastMultiSinger_2021} and DelightfulTTS 2 \cite{liu_DelightfulTTSEndtoEndSpeech_2022}.

\subsection{Normalizing flows}

Although VAE and GAN were frequently utilized in audio generation, neither allowed for an exact evaluation of the probability density of new points \cite{kobyzev_NormalizingFlowsIntroduction_2021}.
Normalizing Flows (NF) are a family of generative models with tractable distributions that enable exact density evaluation and sampling.
The network is made up of two "flows" that move in opposite directions.
One flow starts with a base density, which we call noise, and progresses to a more complex density.
The opposing flow reverses the direction, transforming the complex density back into the base density.
The movement from base to complex is known as the generating direction, whereas the reverse is known as the normalizing direction.
The term normalizing flow refers to the notion that the normalizing direction makes a complex distribution more regular, or normal.
The normal distribution is typically used for base density, which is another reason for the name.
Similar to how we layer transformations in a deep neural network, we compose several simple functions to generate complex behavior.
These functions cannot be chosen arbitrarily because the flow must be in both directions; hence, the functions chosen must be invertible.
Using a characteristic of the invertible function composition, we can create a likelihood-based estimation of the parameters that we can apply to train the model.
In this setup, data generation is simple; utilizing the generative flow, we can input a sample from the base distribution and generate the required complex distribution sample.
It has been formally proven that if you can build an arbitrarily complex transformation, you will be able to generate any distribution under reasonable assumptions \cite{kobyzev_NormalizingFlowsIntroduction_2021}.

Depending on the nature of the function employed in the flow, there can be significant performance differences that impact training or inference time.
Inverse Autoregressive Flows (IAF) are a type of NF model with a specialized function that allows for efficient synthesis.
The transform is based on an autoregressive network, which means that the current output is only determined by the current and previous input values.
The advantage of this transformation is that the generative flow may be computed in parallel, allowing efficient use of resources such as the GPU.
Although IAF networks can be run in parallel during inference, training with maximum likelihood estimation requires sequential processing.
To allow for parallel training, a probability density distillation strategy is used \cite{oord_ParallelWaveNetFast_2018, ping_ClariNetParallelWave_2019}.
In this method, we try to transfer knowledge from an already-trained teacher to a student model.
Parallel WaveNet \cite{oord_ParallelWaveNetFast_2018} and ClariNet \cite{ping_ClariNetParallelWave_2019} are two IAF models that employ this method.
On the other hand, WaveGlow \cite{prenger_WaveglowFlowbasedGenerative_2019}, FloWaveNet \cite{kim_FloWaveNetGenerativeFlow_2019}, and Glow-TTS \cite{kim_GlowTTSGenerativeFlow_2020} all utilize an affine coupling layer.
Because this layer allows for both parallel training and inference, they can avoid the problems associated with the former.
Other efforts that are worth mentioning are WaveNODE \cite{kim_WaveNODEContinuousNormalizing_2020}, which uses continuous normalizing flow, and WaveFlow \cite{ping_WaveFlowCompactFlowbased_2020}, which uses a dilated 2-D convolutional architecture.

At the end of this section, we'd like to address a problem that can arise when using flow-based networks with audio data.
Because audio is digitally stored in a discrete representation, training a continuous density model on discrete data might lead to poor model performance.
\Textcite{yoon_AudioDequantizationHigh_2020} presented audio dequantization methods that can be deployed in flow-based networks and improved audio generating quality.

\subsection{Transformer networks}

The majority of the material discussed in this section comes from \textcite{zhang_DiveDeepLearning_2023}.

Before we can discuss transformers, we need to talk about attention.
Attention has three components: a query, keys, and values.
We have a database of (key, value) pairs, and our goal is to locate all values that closely match their key with the query.
In the transformer, we improve on this concept by introducing a new type of attention known as self-attention.
Self-attention includes three new functions that accept input and have learnable parameters.
The functions change the input into one of the three previously mentioned components: query, key, and value.
The prefix self refers to the fact that we utilize the same input for both the database and the query.
If we were to translate a sentence, we would expect the translation of the nth word to be determined not only by itself but also by the other words in the sentence.
For this example, the query would be the nth word and the database would be the sentence itself; if the parameters were learned correctly, we would expect to see relevant values for translation as the output of self-attention.
To boost the model's capacity to capture both short- and long-term dependencies, we can concatenate numerous self-attention modules, each with its own set of parameters, resulting in multi-head self-attention.
Furthermore, if we want to prevent the model from attending to future entries, we can use masked multi-head self-attention, which employs a mask to specify which future entries we wish to ignore.

The second essential element of the transformer is positional encoding.
Instead of processing a sequence one at a time, the self-attention in the transformer provides parallel computing.
The consequence is that the sequence's order is not preserved.
The prevailing method for preserving order information is to feed the model with additional input associated with each token.
These additional inputs are known as positional encodings.
Position encoding can be absolute or relative, and it can be predefined or learned during training.

At last, the transformer, like an auto-encoder, has both an encoder and a decoder.
Both the encoder and decoder are made up of a stack of identical layers, each with two sublayers.
The first sublayer is multi-head self-attention, whereas the second is a feed-forward network.
In addition, each identical layer contains a residual connection surrounding both sublayers that follows layer normalization.
Unlike the encoder, the decoder employs both encoder-decoder attention and masked multi-head self-attention at the input.
The encoder-decoder attention is a normal multi-head attention with queries from the decoder and (key, value) pairs from the encoder.
Before the input is fed into the network, positional embedding is applied.

Transformers were primarily intended for natural language processing but were then used for images with the vision transformer and lately for audio signals \cite{zaman_SurveyAudioClassification_2023}.
They have revolutionized modern deep learning by providing the ability to model long-term sequences \cite{verma_GenerativeModelRaw_2021}.
On the other hand, transformers are generally referred to as data-hungry since they require a large amount of training data \cite{zaman_SurveyAudioClassification_2023}.
The attention mechanism's quadratic complexity makes it difficult to process long sequences \cite{verma_GenerativeModelRaw_2021}.
To use transformers with audio, we would convert signals into visual spectrograms and divide them into "patches" that are then treated as separate input tokens, analogous to text \cite{zaman_SurveyAudioClassification_2023}.
There are many works that use the transformer architecture, including Music Transformer \cite{huang_MusicTransformer_2018}, FastSpeech \cite{ren_FastSpeechFastRobust_2019}, Wave2Midi2Wave \cite{hawthorne_EnablingFactorizedPiano_2019}, \textcite{li_NeuralSpeechSynthesis_2019}, RobuTrans \cite{li_RobuTransRobustTransformerBased_2020}, Jukebox \cite{dhariwal_JukeboxGenerativeModel_2020}, AlignTTS \cite{zeng_AlignttsEfficientFeedForward_2020}, Multi-Track Music Machine \cite{ens_MMMExploringConditional_2020}, JDI-T \cite{lim_JDITJointlyTrained_2020}, AdaSpeech \cite{chen_AdaSpeechAdaptiveText_2021}, FastPitch \cite{lancucki_FastpitchParallelTexttoSpeech_2021}, \textcite{verma_GenerativeModelRaw_2021}, Controllable Music Transformer \cite{di_VideoBackgroundMusic_2021}, \textcite{lee_DirectSpeechtospeechTranslation_2022a}, SpeechT5 \cite{ao_SpeechT5UnifiedModalEncoderDecoder_2022}, CPS \cite{wang_CPSFullSongStyleConditioned_2022}, FastSpeech 2 \cite{ren_FastSpeechFastHighQuality_2022}, FIGARO \cite{rutte_FIGAROControllableMusic_2022}, HAT \cite{zhang_StructureEnhancedPopMusic_2022}, ELMG \cite{bao_GeneratingMusicEmotions_2023}, AudioLM \cite{borsos_AudioLMLanguageModeling_2023}, VALL-E \cite{wang_NeuralCodecLanguage_2023}, MusicLM \cite{agostinelli_MusicLMGeneratingMusic_2023}, SingSong \cite{donahue_SingSongGeneratingMusical_2023}, SPEAR-TTS \cite{kharitonov_SpeakReadPrompt_2023}, AudioGen \cite{kreuk_AudioGenTextuallyGuided_2023}, VALL-E X \cite{zhang_SpeakForeignLanguages_2023}, dGSLM \cite{nguyen_GenerativeSpokenDialogue_2023}, VioLA \cite{wang_VioLAUnifiedCodec_2023}, MuseCoco \cite{lu_MuseCocoGeneratingSymbolic_2023}, Im2Wav \cite{sheffer_HearYourTrue_2023}, AudioPaLM \cite{rubenstein_AudioPaLMLargeLanguage_2023}, VampNet \cite{garcia_VampNetMusicGeneration_2023}, LM-VC \cite{wang_LMVCZeroshotVoice_2023}, UniAudio \cite{yang_UniAudioAudioFoundation_2023}, and MusicGen \cite{copet_SimpleControllableMusic_2023}.

LakhNES \cite{donahue_LakhNESImprovingMultiinstrumental_2019} and REMI \cite{huang_PopMusicTransformer_2020} use Transformer-XL, an extension of the Transformer that can, in theory, encode arbitrary long contexts into fixed-length representations.
This is accomplished by providing a recurrence mechanism \cite{dai_TransformerXLAttentiveLanguage_2019}, wherein the preceding segment is cached for later usage as an expanded context for the subsequent segment.
Furthermore, to support the recurrence mechanism, it introduces an expanded positional encoding scheme known as relative positional encoding, which keeps positional information coherent when reusing states.
In addition to Transformer-XL, \textcite{hawthorne_GeneralpurposeLongcontextAutoregressive_2022} and \textcite{yu_MuseformerTransformerFine_2022} presented Perceiver AR and Museformer as alternatives to tackle problems that require extended contexts.

Finally, another extension to the transformer has been successful for various speech tasks \cite{bai_AlignmentAwareAcousticText_2022}.
Convolution-augmented Transformer (Conformer) extends the Transformer by incorporating convolution and self-attention between two feed-forward modules; this cascade of modules is a single Conformer block.
It integrates a relative positional encoding scheme, a method adopted from the described Transformer-XL to improve generalization for diverse input lengths \cite{gulati_ConformerConvolutionaugmentedTransformer_2020}.
Papers utilizing the conformer are SpeechNet \cite{chen_SpeechNetUniversalModularized_2021}, \(\text{A}^3\text{T}\) \cite{bai_AlignmentAwareAcousticText_2022}, VQTTS \cite{du_VQTTSHighFidelityTexttoSpeech_2022}, \textcite{popuri_EnhancedDirectSpeechtoSpeech_2022}, and SoundStorm \cite{borsos_SoundStormEfficientParallel_2023}.

\subsection{Diffusion models}

Diffusion models are generative models inspired by non-equilibrium thermodynamics \cite{mehrish_ReviewDeepLearning_2023}.
Diffusion models, like normalizing flows, consist of two processes: forward and reverse.
The forward process converts the data to a conventional Gaussian distribution by constructing a Markov chain of diffusion steps with a predetermined noise schedule.
The reverse method gradually reconstructs data samples from the noise using an interference noise schedule.
Unlike other architectures that change the distribution of data, such as variational auto-encoders and normalizing flows, diffusion models maintain the dimensionality of the latent variables fixed.
Because the dimensionality of the latent variables must be fixed during the iterative generation process, which can result in slow inference speed in high-dimensional spaces \cite{liu_AudioLDMTexttoAudioGeneration_2023}.
A potential solution is to utilize a more compressed representation, such as a mel-spectrogram, instead of a short-time Fourier transform.
Papers employing diffusion models include WaveGrad \cite{chen_WaveGradEstimatingGradients_2020}, DiffWave \cite{kong_DiffWaveVersatileDiffusion_2021}, Diff-TTS \cite{jeong_DiffTTSDenoisingDiffusion_2021}, Grad-TTS \cite{popov_GradTTSDiffusionProbabilistic_2021}, DiffuSE \cite{lu_StudySpeechEnhancement_2021}, FastDiff \cite{huang_FastDiffFastConditional_2022}, CDiffuSE \cite{lu_ConditionalDiffusionProbabilistic_2022}, Guided-TTS \cite{kim_GuidedTTSDiffusionModel_2022a}, Guided-TTS 2 \cite{kim_GuidedTTSDiffusionModel_2022}, DiffSinger \cite{liu_DiffSingerSingingVoice_2022}, UNIVERSE \cite{serra_UniversalSpeechEnhancement_2022}, Diffsound \cite{yang_DiffsoundDiscreteDiffusion_2023}, Noise2Music \cite{huang_Noise2MusicTextconditionedMusic_2023}, DiffAVA \cite{mo_DiffAVAPersonalizedTexttoAudio_2023}, MeLoDy \cite{lam_EfficientNeuralMusic_2023}, Tango \cite{ghosal_TexttoAudioGenerationUsing_2023}, SRTNet \cite{qiu_SRTNETTimeDomain_2023}, MusicLDM \cite{chen_MusicLDMEnhancingNovelty_2023}, JEN-1 \cite{li_JEN1TextGuidedUniversal_2023}, AudioLDM \cite{liu_AudioLDMTexttoAudioGeneration_2023}, \textcite{bai_AcceleratingDiffusionBasedTexttoAudio_2023}, ERNIE-Music \cite{zhu_ERNIEMusicTexttoWaveformMusic_2023}, and Re-AudioLDM \cite{yuan_RetrievalAugmentedTexttoAudioGeneration_2024}.

Transformer and diffusion models were the most popular designs discussed in this article.
As a result, in the final half of this chapter, we will list some works that use both the transformer and diffusion models.
These works include \textcite{hawthorne_MultiinstrumentMusicSynthesis_2022}, Make-An-Audio 2 \cite{huang_MakeAnAudioTemporalEnhancedTexttoAudio_2023}, NaturalSpeech 2 \cite{shen_NaturalSpeechLatentDiffusion_2023}, Grad-StyleSpeech \cite{kang_GradStyleSpeechAnySpeakerAdaptive_2023}, Make-An-Audio \cite{huang_MakeAnAudioTextToAudioGeneration_2023}, AudioLDM 2 \cite{liu_AudioLDMLearningHolistic_2023}, and Moûsai \cite{schneider_MoUsaiTexttoMusic_2023}.

\section{Evaluation metrics}
\label{sec:eval_metrics}

Evaluations could be considered the most important piece of the puzzle.
By introducing evaluations, we are able to quantify progress.
We can compare, improve, and optimize our models, all thanks to evaluation metrics.
We will not mention domain-specific evaluation metrics such as the character error rate used in text-to-speech\nocite{kang_GradStyleSpeechAnySpeakerAdaptive_2023} or the perceptual evaluation of speech quality used in speech enhancement\nocite{qiu_SRTNETTimeDomain_2023}.
There are many more widely used metrics that we will not mention in the following sections. Some of them are: Nearest neighbor comparisons, Number of statistically-different bins, Kernel Inception Distance, and CLIP score.

\subsection{Human evaluation}

As humans, we are constantly exposed to various sorts of sounds, which provides us with a wealth of expertise when attempting to distinguish between real and manufactured audio.
The audio-generating method we are seeking to construct is intended to trick people into thinking the sound is a recording rather than synthesis.
As a result, who better to judge the success of such systems than the ones we're attempting to fool?
Human evaluation is the gold standard for assessing audio quality. The human ear is particularly sensitive to irregularities, which are disruptive for the listener \cite{engel_GANSynthAdversarialNeural_2019}.
Intuitively, it is simple to label an audio sample as good or bad, real or fake, but it is much more challenging to document a procedure derived from our thinking that may be used to evaluate future audio samples.
The human assessment is often carried out with a defined number of listeners who listen and rate the audio on a 1-5 Likert scale. This type of test is termed the Mean-Opinion Score (MOS) \cite{maiti_SpeechlmscoreEvaluatingSpeech_2023}.
While MOS is used to evaluate naturalness, similarity MOS is used to assess how similar the generated and real samples are \cite{arik_NeuralVoiceCloning_2018}.
Another metric, known as the comparative MOS, can be used to compare two systems by subtracting their MOS values.
We may also calculate it by providing listeners with two audio samples generated by different models and immediately judging which one is better \cite{li_NeuralSpeechSynthesis_2019}.
\Textcite{oord_ParallelWaveNetFast_2018} discovered that preference scores from a paired comparison test, frequently referred to as the A/B test, were more reliable than the MOS score.
Many alternative human evaluation metrics have been proposed for music; domain-specific metrics include melody, groove, consonance, coherence, and integrity.
The biggest disadvantage of human evaluation is that the findings cannot be replicated exactly. This means that the concrete numbers in the evaluation are unimportant, and only the link between them is crucial.
This stems from the inherent subjectivity of human evaluation as well as biases or predispositions for specific auditory features.

\subsection{Inception score}

The Inception Score (IS) is a perceptual metric that correlates with human evaluation.
The Inception score is calculated by applying a pretrained Inception classifier to the generative model's output.
The IS is defined as the mean Kullback-Leibler divergence between the conditional output class label distribution and the labels' marginal distribution.
It evaluates the diversity and quality of the audio outputs and prefers generated samples that the model can confidently classify.
With \(N\) samples, the measure ranges from \(1\) to \(N\).
The IS is maximized when the classifier is confident in every classification of the generated sample and each label is predicted equally often \cite{donahue_AdversarialAudioSynthesis_2019}.
Normally, the Inception classifier is trained using the ImageNet dataset, which may not be compatible with audio spectrograms.
This will cause the classifier to be unable to separate the data into meaningful categories, resulting in a low IS score.
An extension to the IS called Modified Inception Score (mIS) measures the within-class diversity of samples in addition to the IS which favors sharp and clear samples \cite{kong_DiffWaveVersatileDiffusion_2021}.

\subsection{Fréchet distance}

The inception score is based solely on the generated samples, not taking into consideration the real samples \cite{haque_HighFidelityAudioGeneration_2020}.
The Fréchet Inception Distance (FID) score addresses this issue by comparing the generated samples with the real ones.
The FID calculates the Fréchet distance between two distributions for both generated and real samples using distribution parameters taken from the intermediate layer of the pretrained Inception Network \cite{haque_HighFidelityAudioGeneration_2020}.
The lower the FID score, the higher the perceived generation quality.
It is frequently used to assess the fidelity of generated samples in the image generation domain \cite{yang_DiffsoundDiscreteDiffusion_2023}.
This metric was found to correlate with perceptual quality and diversity on synthetic distributions \cite{engel_GANSynthAdversarialNeural_2019}.
The Inception Network is trained on the ImageNet dataset, which is purpose-built for images, but this does not ensure it will function for spectrograms. It may be unable to classify the spectrograms into any meaningful categories, resulting in unsatisfactory results \cite{haque_GuidedGenerativeAdversarial_2021}.

Fréchet Audio Distance (FAD) is a perceptual metric adapted from the FID for the audio domain. Unlike reference-based metrics, FAD measures the distance between the generated audio distribution and the real audio distribution using a pretrained audio classifier that does not use reference audio samples.
The VGGish model \cite{hershey_CNNArchitecturesLargescale_2017} is used to extract the characteristics of both generated and real audio \cite{liu_AudioLDMLearningHolistic_2023}.
As with the FID, the lower the score, the better the audio fidelity.
According to \textcite{kreuk_AudioGenTextuallyGuided_2023}, the FAD correlates well with human perceptions of audio quality.
The FAD was found to be robust against noise, computationally efficient, consistent with human judgments, and sensitive to intra-class mode dropping \cite{nistal_ComparingRepresentationsAudio_2021}.
Although FAD may indicate good audio quality, it does not necessarily indicate that the sample is desired or relevant \cite{agostinelli_MusicLMGeneratingMusic_2023}.
For instance, in text-to-speech applications, low-FAD audio may be generated that does not match the input text.
According to \textcite{binkowski_HighFidelitySpeech_2019}, the FAD measure is not appropriate for evaluating text-to-speech models since it was created for music.
While according to \textcite{zhu_ERNIEMusicTexttoWaveformMusic_2023}, calculating the similarity between real and generated samples does not account for sample quality.
Another similar metric called the Fréchet DeepSpeech Distance (FDSD) also uses the Fréchet distance on audio features extracted by a speech recognition model \cite{natsiou_AudioRepresentationsDeep_2021}. \Textcite{donahue_EndtoEndAdversarialTexttoSpeech_2021} found the FDSD to be unreliable in their use case.

The last Fréchet metric that is important to discuss is the Fréchet Distance (FD). Unlike the FAD, which extracts features using the VGGish \cite{hershey_CNNArchitecturesLargescale_2017} model, FD employs the PANN \cite{kong_PANNsLargeScalePretrained_2020} model.
The model change enables the FD to use different audio representations as input. PANN \cite{kong_PANNsLargeScalePretrained_2020} uses mel-spectrogram as input, whereas VGGish \cite{hershey_CNNArchitecturesLargescale_2017} uses raw waveform.
FD evaluates audio quality using an audio embedding model to measure the similarity between the embedding space of generations and that of targets \cite{chen_MusicLDMEnhancingNovelty_2023}.

\subsection{Kullback-Leibler divergence}

Kullback-Leibler (KL) divergence is a reference-dependent metric that computes the divergence between the generated and reference audio distributions.
It uses a pretrained classifier to obtain the probabilities of generated and reference samples and then calculates the KL divergence between the distributions.
The probabilities are computed over the class predictions of the pretrained classifier.
A low KL divergence score may indicate that a generated audio sample shares concepts with the given reference \cite{li_JEN1TextGuidedUniversal_2023}.
In music, this could indicate that the created audio has similar acoustic characteristics \cite{agostinelli_MusicLMGeneratingMusic_2023}.
While the FAD measure is more related to human perception \cite{ghosal_TexttoAudioGenerationUsing_2023}, the KL measure reflects more on the broader audio concepts occurring in the sample \cite{kreuk_AudioGenTextuallyGuided_2023}.

\section{Conclusion}

The development of deep learning methods has significantly changed the field of audio generation.
In this work, we have presented three important parts of building a deep learning model for the task of audio generation.
For audio representation, we have presented two long-standing champions and a third up-and-comer. We explained five architectures and listed work that implements them in the field of audio generation. Finally, we presented the four most common evaluations in the works we examined.
While the first three architectures mentioned above seem to have lost importance in recent years, the transformer and diffusion models seem to have taken their place.
This could be due to the popularization of large language models such as ChatGPT or, in the case of the diffusion models, diffusion-based text-to-image generators.

With the ever-increasing computing power and availability of large databases, it looks like the age of deep learning has only just begun.
Just as deep learning has allowed us to move from domain-dependent features and methods to a more universal solution, more recent work has attempted to move from a single task or purpose to a multi-task or even multi-modality model.

\printbibliography

\end{document}